\documentclass[twocolumn,aps,pra,letterpaper]{revtex4-1}
\usepackage[pass]{geometry}

\usepackage[pdftex]{graphicx,color}
\begin{document}
%
\title{Phase Analysis for Frequency Standards in the Microwave and Optical Domains} %
%
%
%

\author{M.~Kazda, V.~Gerginov, N.~Huntemann, B.~Lipphardt, and~S.~Weyers}
\affiliation{Physikalisch-Technische Bundesanstalt, Bundesallee 100, 38116 Braunschweig, Germany}

\begin{abstract}


Coherent manipulation of atomic states is a key concept in high-precision spectroscopy and used in atomic fountain clocks and a number of optical frequency standards. Operation of these standards can involve a number of cyclic switching processes, which may induce cycle synchronous phase excursions of the interrogation signal and thus lead to shifts in the output of the frequency standard. We have built a FPGA-based phase analyzer to investigate these effects and conducted measurements on two frequency standards. For the caesium fountain PTB-CSF2 we were able to exclude phase variations of the microwave source at the level of a few $\mu$rad, corresponding to relative frequency shifts of less than 10$^{-16}$. In the optical domain, we investigated phase variations in PTB's Yb$^+$ optical frequency standard and made detailed measurements of AOM chirps and their scaling with duty cycle and driving power. We ascertained that cycle-synchronous as well as long-term phase excursion do not cause frequency shifts larger than 10$^{-18}$. 

\end{abstract}

\maketitle


%


\section{Introduction}
\label{sec_intro}




In atomic frequency standards the utilization of separated oscillatory fields (Ramsey method) is well established because of its increased spectral resolution compared to the utilization of a single oscillatory field for the same interaction time and because of other advantages \cite{Ramsey1990,Yudin2010}. However, the pulsed operation mode of fountain clocks and optical frequency standards involves a number of cyclic switching processes, which may cause cycle synchronous phase differences of the oscillatory field in the two short Ramsey interaction periods and thereby induce shifts in the output frequency. In a fountain clock, the origin of detrimental cyclic phase excursions can be the utilization of an interferometric microwave switch \cite{Santarelli2009}, employed for switching off the microwave signal between the two subsequent atom-field interactions. Such cycle synchronous phase variations at the few $\mu$rad level result in frequency shifts in the range of 10$^{-16}$, which is comparable to the level of the dominant uncertainty contributions in present-day fountain clocks like PTB-CSF2 \cite{Gerginov2010,Weyers2012}.

To detect small phase variations of the microwave field in fountain clocks, sophisticated phase analyzing techniques have been developed \cite{Santarelli2006,Santarelli2009}, capable of coherent averaging synchronous to the fountain cycle. Based on these techniques we have developed a phase analyzer system utilizing a commercially available field programmable gate array (FPGA) with an analog-digital converter front-end. The data acquisition and filtering is entirely carried out within the FPGA, enabling the use of variable filters and decimation techniques as well as digital I/Q demodulation and phase detection. Due to the large on-board memory, high-resolution measurements can be performed for long interrogation cycles. For our caesium fountain clocks, after 12 hours of averaging, the analyzer allows for the detection of cycle-synchronous phase excursions that would cause frequency shifts of less than 10$^{-16}$ .

The increased processing speed of our system enables likewise its application in optical frequency standards. In our single-ion optical frequency standard based on the $^2S_{1/2}\rightarrow{}^2F_{7/2}$ electric octupole transition of $^{171}$Yb$^+$ a Ramsey-type excitation scheme is employed to avoid interaction induced frequency shifts \cite{Huntemann2012}. Because in this frequency standard there is no active stabilization of the optical path length between the interrogation laser and the position of the ion, cycle synchronous phase excursions can lead to an error in the observed transition frequency. To investigate the magnitude of these phase excursions, we set up an interferometer and modified our phase analyzer for the analysis of the beat signals. After 24 hours of data acquisition, we are able to determine phase excursions that correspond to frequency shifts of less than 10$^{-18}$.

In the following we will first give an overview of the measurement system and the techniques common to both measurement cases in the microwave and optical domains. Then we will explain the special requirements for measurements at the atomic fountain clocks and present results. The Yb$^+$-interrogation-cycle will be introduced and the setup required for optical phase measurements will be defined. With this setup, we measured the phase excursion due to cyclic changes in the power of the signal that drives an acousto-optic modulator (AOM), known as the AOM chirp.

\section{Design}
\label{sec_design}

\begin{figure*}
  \includegraphics[width=\textwidth]{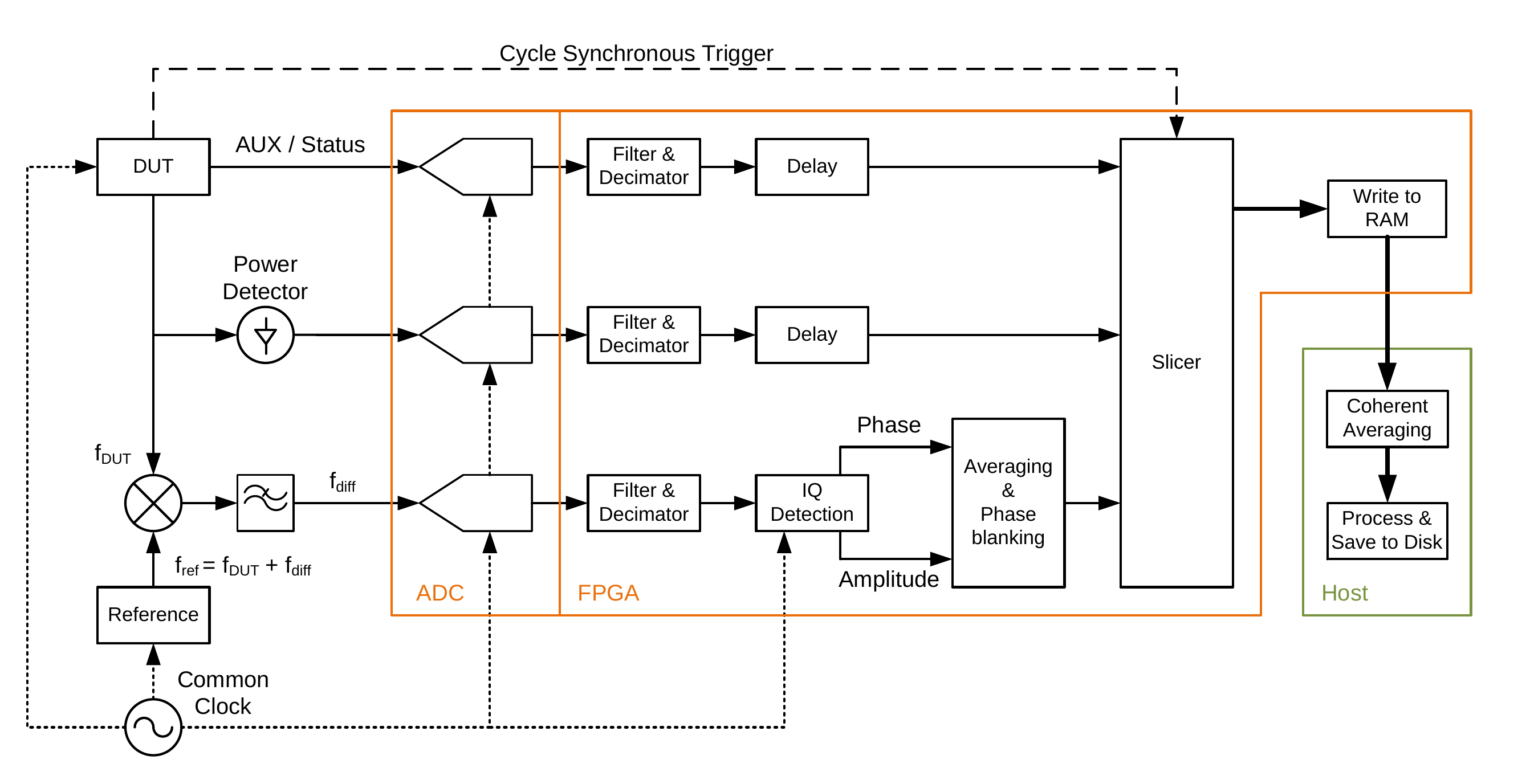}
	\caption{Analyzer setup consisting of analog front-end, AD-converter and digital data processing.}
	\label{fig_FPGA} 
\end{figure*}

In cyclic processes, signal transients buried deeply in uncorrelated noise can be recovered by coherent averaging. To investigate such phase transients in our atomic frequency standards, we have built an analyzer (Fig.~\ref{fig_FPGA}) based on this averaging technique. It supports variable analysis bandwidths and cycle lengths up to several seconds with a time resolution of 1~$\mu$s. Possible sources of phase variations, e.g., temperature changes, can be correlated by using two additional channels that are aligned with the averaged phase data.

Since the frequencies $f_{DUT}$ of the devices under test (DUT) investigated here are outside the limited bandwidth of the system, they are downmixed to a frequency $f_{diff}$. A reference synthesizer is used to generate the required frequency $f_{ref} = f_{DUT} +f_{diff}$. After downmixing, the signal is fed to an anti-aliasing filter and then AD-converted with sample rate of 120~MS/sec. Once the signal is digitized, all data processing is performed in the FPGA. First, the signal is filtered and decimated \cite{Crochiere1975}. This stage allows for variable decimation factors, depending on the required time resolution. The time resolution, decimation factor and difference frequency $f_{diff}$ are chosen according to specific requirements. The decimated signal is fed to an IQ-detector algorithm, where a direct digital synthesis (DDS) is implemented in software to generate a sinusoidal signal and the corresponding quadrature signal. Both signals are mixed with the decimated signal and phase and amplitude are calculated. Depending on the application, phase variations larger than 2$\pi$ during one measurement cycle are possible. We therefore need to implement a phase unwrapping algorithm in a later stage. However, when the signal level from the DUT is too low, the IQ-detector will give almost random phase values, which cannot be processed reliably by an unwrapping algorithm. Therefore, the amplitude information is used to mark phase values as valid or invalid. The output of this averaging \& phase blanking block is a continuous data stream.

To facilitate coherent averaging, the stream has to be divided in packages synchronous to the DUT cycle. A cycle synchronous trigger is fed to a digital input of the FPGA and the data stream is divided into evenly sized packages.  
These data packages are written to  RAM connected directly to the FPGA. The host monitors the acquisition and reads completed slices into the host memory, where they are coherently averaged. It also performs phase unwrapping, various statistical calculations and consistency checks. The averaged data as well as statistics are then saved to disk.


Parallel to the phase detection, the power level of the DUT and temperature or status information from the DUT are monitored. These signals are AD-converted, filtered and decimated, similar to the phase signal. To compensate for delays in the IQ-demodulation of the phase signal, a small delay needs to be inserted. The phase stream and the other data streams are now aligned and can be averaged.

The analyzer is conceived for detecting periodic phase perturbations. A common clock is employed to ensure a stable phase relation between the analog signals and the digital detection and  demodulation stages. For data acquisition and processing, a PXI System from National Instruments in combination with a 7966R FPGA card and a 5734R ADC converter front-end is used.



\section{Investigation of the Caesium Fountain}
\label{sec_results_fountain}



Atomic fountain clocks use the Ramsey interrogation method to provide a realization of the SI second  \cite{Wynands2005a}. They employ a cyclic process, consisting of multiple stages. In CSF2 \cite{Gerginov2010}, caesium atoms are first cooled in the intersection region of six laser beams and subsequently launched vertically by laser frequency detuning. After launching, all laser beams are switched off and the cloud follows a ballistic trajectory. In a state selection microwave cavity the atoms are prepared in a single hyperfine magnetic substate, before on the way up and on the way down the atoms pass through the Ramsey cavity where they interact with the microwave field. After the second cavity passage, detection lasers probe the atom state population. At each cycle the frequency of the microwave field in the Ramsey cavity is detuned to probe the transition probabilities at both sides of the central Ramsey fringe alternately. The resulting discriminator signal is used by a computer to control the output frequency $f_{DUT}$ of the microwave synthesizer. 

Because of the resulting cyclic changes of $f_{DUT}$, the reference frequency $f_{ref}$ for the phase analyzer (Fig.~\ref{fig_FPGA}) needs to be adjusted accordingly to make $f_{diff}$ constant. The required measurement time to evaluate phase excursions is significantly reduced by referencing the microwave synthesizer of the fountain to an optically stabilized 9.6~GHz oscillator \cite{Weyers2009, Tamm2014}. This oscillator exhibits a very low phase noise of about 100 $\mu$rad rms (single cycle). It is used as the common clock and divided down to 10 MHz to lock the FPGA and ADC clocks. 

\begin{figure}
  \includegraphics[width=\columnwidth]{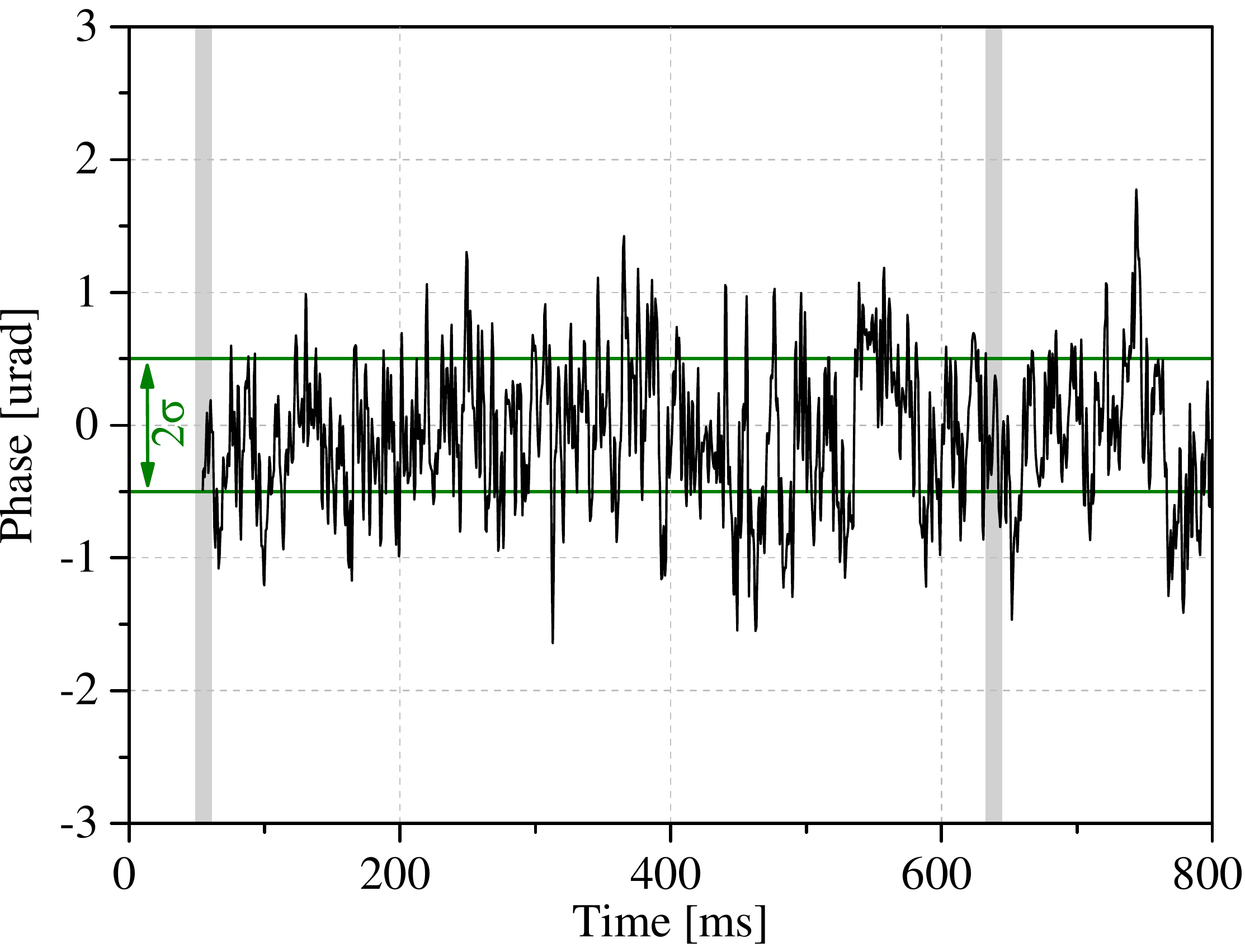}
 	\caption{Phase measurement of the microwave interrogation signal at CSF2 during and after the two Ramsey interactions (marked by the gray shaded stripes). The indicated 2$\sigma$ interval is calculated from the independently characterized DUT noise and is in good agreement with the standard deviation of the measured phase values.}	
	\label{fig_Meas_CSF2} 
\end{figure}

In Fig.~\ref{fig_Meas_CSF2}, a phase measurement of the Ramsey interrogation signal at PTB's fountain clock CSF2 with a time resolution of 0.5~ms is shown. The data was averaged for about 12~h, corresponding to 35000 fountain cycles. The two gray shaded stripes mark the occurrence of the two successive Ramsey interactions, each with a duration of about 10~ms. From the data in Fig.~\ref{fig_Meas_CSF2} we estimate the phase difference between the two interactions to be smaller than 1 $\mu$rad, corresponding to a frequency shift of less than $3\times10^{-17}$. The standard deviation of the data set is 0.55 $\mu$rad.

\section{Investigation of the Yb$^+$ optical frequency standard}
\label{sec_results_yb}

At PTB we are currently investigating optical frequency standards based on the $^{171}$Yb$^+$ quadrupole and octupole transitions \cite{Huntemann2012a, Tamm2014}. For the octupole transition we employ the Hyper-Ramsey-scheme (HRS) \cite{Yudin2010, Huntemann2012}. In this scheme, phase and frequency steps are applied to the interrogation laser light via an AOM controlled by a DDS. Changes in the power of the signal that drives the modulator lead to temperature changes of its crystal and in turn change the corresponding optical path length due to thermal expansion and the temperature dependance of the refractive index. The magnitude of the so-called AOM chirp depends on the power and the duty cycle of the driving signal~\cite{Degenhardt2005, Rosenband2008}. To reduce this effect, mechanical shutters are used to block the light during dark times and the AOM driving power is only switched off for short intervals in order to define the edges of the Ramsey pulses and during phase changes of the DDS. 

So far, no active path length stabilization is implemented for the free-space probe laser beam. Residual phase fluctuations and drifts can therefore introduce frequency shifts. These can not only result from afore mentioned AOM chirps but also from air turbulences or thermal effects in optical parts or the mechanical setup. However, effects that are synchronous with the interrogation cycle require particularly careful consideration.

\begin{figure}
  \includegraphics[width=\columnwidth]{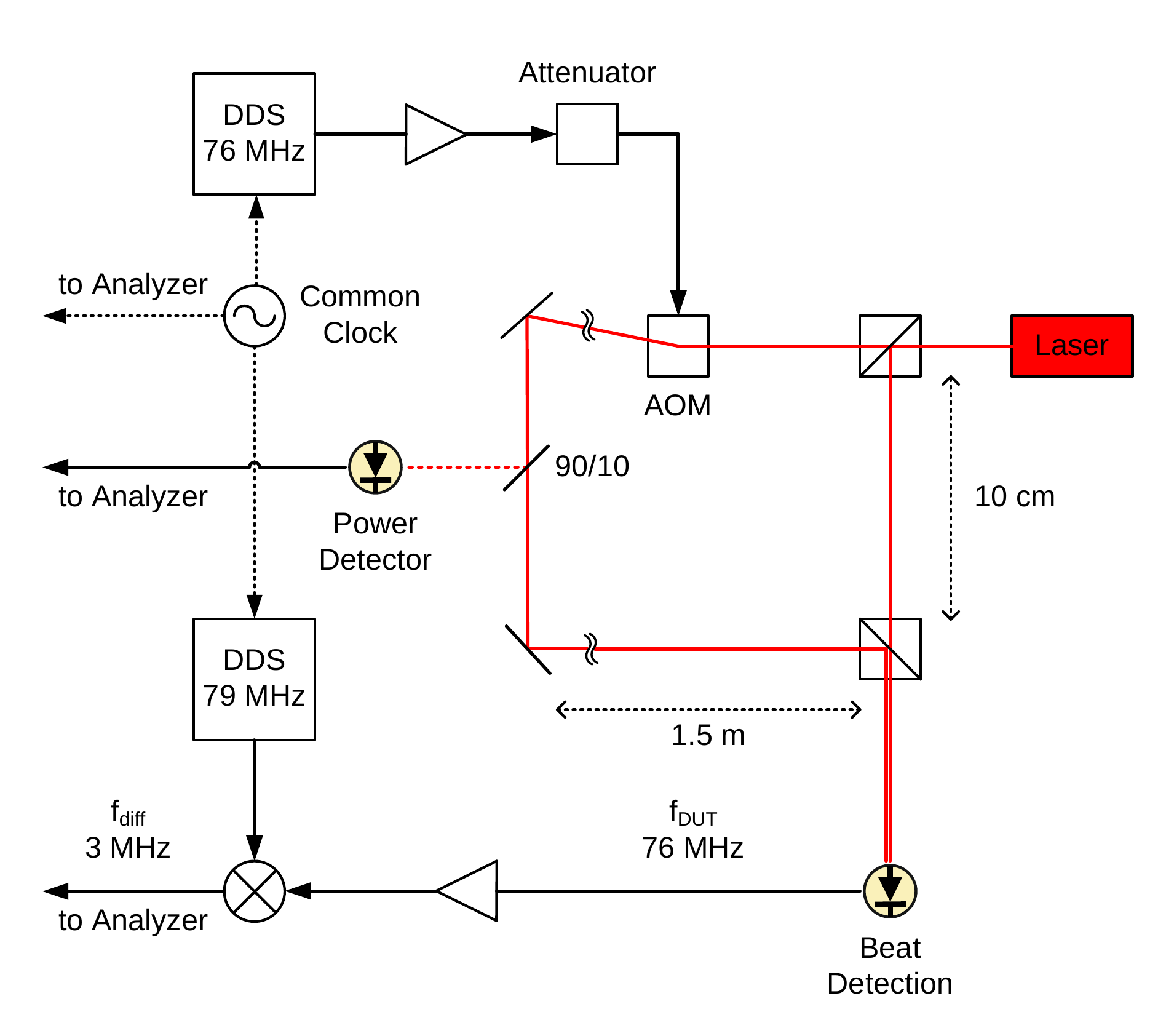}
	\caption{Interferometric phase analysis setup at the Yb$^+$ optical frequency standard. For details see text.}
	\label{fig_opt_table}
\end{figure}

To evaluate such shift effects for the optical frequency standard, we need to determine the phase of the laser light after passage through the AOM and free space propagation by setting up an interferometer (see Fig.~\ref{fig_opt_table}). In the long arm of the interferometer the light is diffracted by a TeO$_2$-AOM and a small fraction of the laser light is used to detect the optical power. By beating the light from the long arm against the light from the short reference arm, we detect the optical phase. 

A 76~MHz-signal generated by a DDS is used to drive the AOM, resulting in a 76~MHz beat note on the photo-detector ($f_{DUT}$). This beat note is amplified and mixed with 79~MHz generated by a second DDS. The resulting 3~MHz-signal ($f_{diff}$) is fed directly to the phase analyzer. To correct for phase drifts and excursions resulting from the DDS that drives the AOM, we measured the 76~MHz DDS-signal directly and subtracted the obtained phase data from the beat note measurements. Thermally induced phase changes in the photo-detector used for beat detection are not corrected for in the results presented here.

\begin{figure}
  \includegraphics[width=\columnwidth]{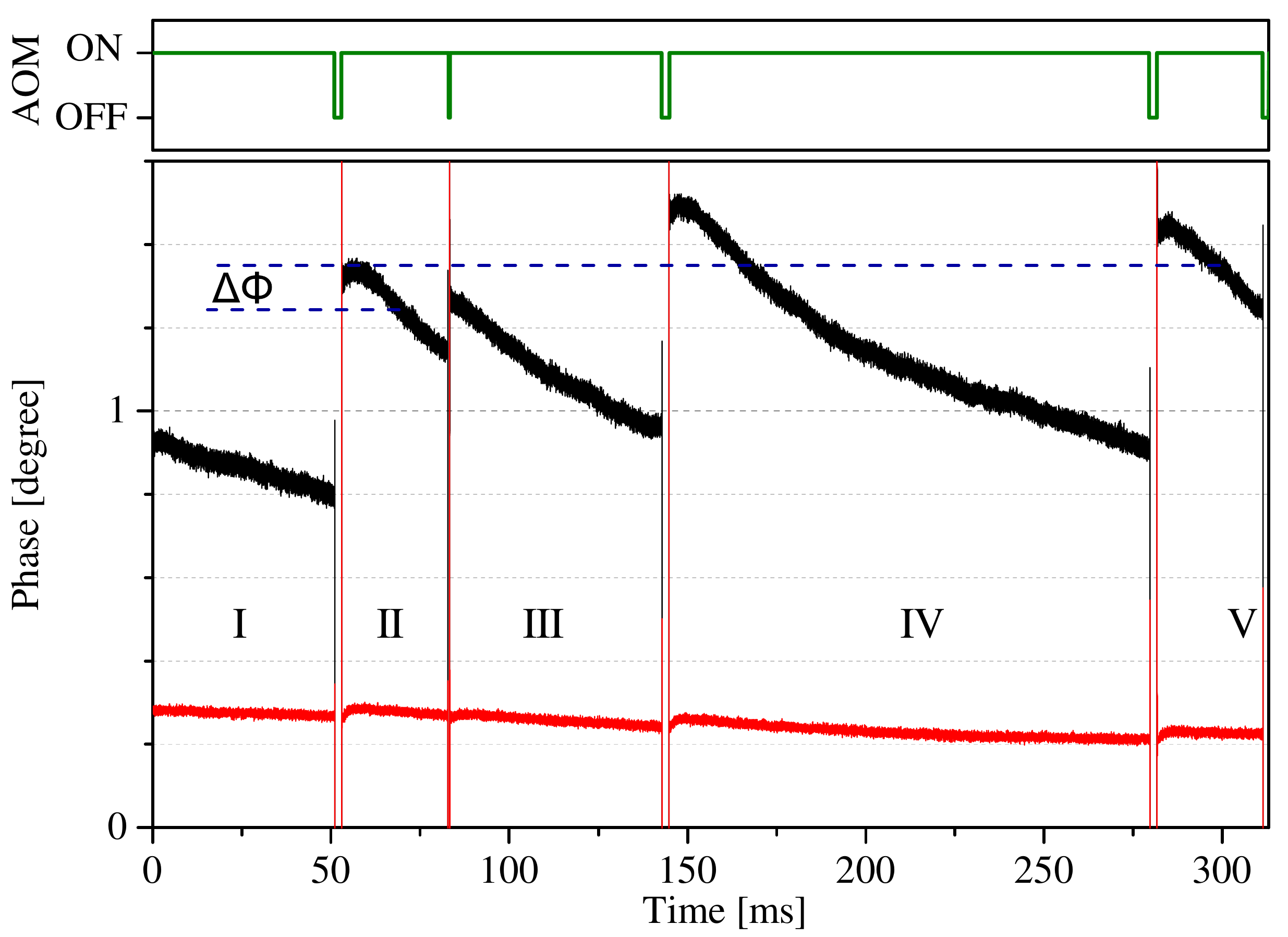}
	\caption{
Optical phase excursions averaged over repeated measurement cycles. The black curve was acquired for an AOM driving power of 1~W, for the red curve the nominal driving power of 50~mW was used. As illustrated in the upper part of the figure, the AOM is switched off for either 0.5 or 2~ms. Roman numbers mark the different phases of the interrogation cycle (see text). The observed phase difference $\Delta\Phi=0.1$~degree between the two Ramsey interactions in phases II and V is attributed to underlying long-term thermal drifts of the interferometer setup and should therefore be independent of the AOM driving power. Please note that there is a gap in the data between phases V and I.
					}
  \label{fig_Meas_Yb}	
\end{figure}

In contrast to the standard Ramsey method mentioned earlier, the Hyper-Ramsey-scheme contains an additional pulse between the two Ramsey-interactions \cite{Yudin2010, Huntemann2012}. The particular scheme investigated here consists of five phases, separated by short AOM off gaps with a duration of 2~ms or 0.5~ms as illustrated in Fig.~\ref{fig_Meas_Yb}. In phase I readout and state preparation are performed and phases II and V are the actual Ramsey interactions. The additional pulse occurs in phase III and phase IV indicates the free precession time.

After sufficient cycle-synchronous averaging, various mechanisms affecting the phase can be identified, among them particularly AOM chirps. To accentuate their impact, we measured the phase excursions with an increased driving power of 1~W and for comparison under normal conditions with a driving power of 50~mW (shown in black and red respectively in Fig.~\ref{fig_Meas_Yb}). Distinctive phase jumps during the off gaps can be traced back to the AOM crystal cooling down. Their magnitude depends on the AOM driving power and the switch-off time. At high driving power these phase jumps are 0.5~degree and 0.1~degree for off-times of 2~ms and 0.5~ms, respectively. Under normal operation conditions with low driving power, the phase jumps are largely reduced and smaller than 0.01~degree. When switched back on, the AOM heats up towards the steady state condition. This leads to phase drifts during the different stages of the scheme that can be translated into frequency offsets. The phase gradient during each of the two Ramsey interactions translates into an effective frequency detuning of about 20~mHz for high driving power and less than 2~mHz for low driving power.

In addition to these AOM chirps, Fig.~\ref{fig_Meas_Yb} reveals an additional underlying phase drift, readily observed as the difference between the average phases of the Ramsey pulses II and V. This can be attributed to remaining thermal effects of the optical setup and will be averaged out if the measurement time is significantly larger than the time scale of temperature fluctuations. For the black curve, 100000 cycles were averaged. The difference of the mean phases in the two Ramsey interactions $\Delta\Phi$=0.1~degree is clearly visible. For the red curve, 230000 cycles were averaged and the phase difference is reduced to 0.05~degree. This phase difference corresponds to a relative frequency shift of about $1\times10^{-18}$ that is significantly smaller than the instability of the frequency standard at the chosen measurement time of one day.

\section{Conclusion and Outlook}

We successfully implemented a phase analyzer for detecting cycle-synchronous phase excursions and illustrated two applications for frequency standards in the microwave and optical domains. For the caesium fountain CSF2 we were able to prove that no significant uncertainty contribution is to be expected from phase variations of the microwave source when a continuous microwave signal is used. In a next step we will use the analyzer to validate our interferometric microwave switch for regular use. 

In the optical domain, we measured AOM chirps and their scaling with duty cycle and driving power with high phase resolution and investigated long term thermal drift effects of PTB's Yb$^+$ optical frequency standard setup. Since the phase analyzer provides high time resolution, we plan to investigate the effects of short term phase fluctuations on the transition frequency.

\section*{Acknowledgment}

The authors would like to thank Giorgio Santarelli for inspiring discussions and Christian Sanner for helpful discussions and careful reading of the manuscript. We acknowledge support by the Braunschweig International Graduate School of Metrology B-IGSM. This work was supported by the European Metrology Research Programme (EMRP) in project SIB04. The EMRP is jointly funded by the EMRP participating countries within EURAMET and the European Union.

\end{document}